\pdfoutput=1
\documentclass[sigconf]{acmart}
\AtBeginDocument{%
  }

\usepackage{subcaption}
\usepackage{float}
\usepackage{multirow}
\usepackage{hyperref,xcolor}
\usepackage{array}
\usepackage{pifont}
\newcommand{\cmark}{\ding{51}}
\newcommand{\xmark}{\ding{55}}

\newcolumntype{P}[1]{>{\centering\arraybackslash}p{#1}}

\usepackage[T1]{fontenc}
\definecolor{verylightgray}{gray}{0.95}
\usepackage{microtype}
\usepackage{xspace}
\newcommand{\model}{\textsc{PODTILE}\xspace}

\newcommand{\squishlist}{
 \begin{list}{$\bullet$}
  { \setlength{\itemsep}{3pt}
     \setlength{\parsep}{1pt}
     \setlength{\topsep}{3pt}
     \setlength{\partopsep}{0pt}
     \setlength{\leftmargin}{1.5em}
     \setlength{\labelwidth}{1em}
     \setlength{\labelsep}{0.5em} } }

\newcommand{\squishend}{
  \end{list}  }
  
\newcommand{\squishnumlist}{
\begin{enumerate}[topsep=3pt, itemsep=3pt]
  { \setlength{\itemsep}{3pt}
     \setlength{\parsep}{1pt}
     \setlength{\topsep}{3pt}
     \setlength{\partopsep}{0pt}
     \setlength{\leftmargin}{1.5em}
     \setlength{\labelwidth}{1em}
     \setlength{\labelsep}{0.5em} } }

\copyrightyear{2024}
\acmYear{2024}
\setcopyright{acmlicensed}\acmConference[CIKM '24]{Proceedings of the 33rd
ACM International Conference on Information and Knowledge
Management}{October 21--25, 2024}{Boise, ID, USA}
\acmBooktitle{Proceedings of the 33rd ACM International Conference on
Information and Knowledge Management (CIKM '24), October 21--25, 2024,
Boise, ID, USA}
\acmDOI{00.0000/0000000.0000000}
\acmISBN{979-8-4007-0436-9/24/10}

\begin{document}

%%
%% The "title" command has an optional parameter,
%% allowing the author to define a "short title" to be used in page headers.
\title{\model: Facilitating Podcast Episode Browsing with Auto-generated Chapters}

%%
%% The "author" command and its associated commands are used to define
%% the authors and their affiliations.
%% Of note is the shared affiliation of the first two authors, and the
%% "authornote" and "authornotemark" commands
%% used to denote shared contribution to the research.
%\author{Azin Ghazimatin and Ekaterina Garmash, Gustavo Penha, 
%Kristen Sheets, Martin Achenbach, Oguz Semerci, Remi Galvez, Marcus Tannenberg, 
%Sahitya Mantravadi, Divya Narayanan, Ofeliya Kalaydzhyan, Doug Cole, 
%Ben Carterette, Ann Clifton, Paul Bennett, Claudia Hauff, Mounia Lalmas}
% \leftheader{Ghazimatin and Garmash et al.}
% \authornote{Both authors contributed equally to this research.}
%\email{trovato@corporation.com}
%\orcid{1234-5678-9012}
%\author{G.K.M. Tobin}
%\authornotemark[1]
%\email{webmaster@marysville-ohio.com}
%\affiliation{%
%  \institution{Spotify}
% \city{Germany, US}
% \state{Ohio}
%  \country{}
%  \postcode{43017-6221}
%}

% ---------------
%---------------


\author{Azin Ghazimatin\footnotemark[1]\footnotemark[2]}
%\orcid{0000-0002-0282-2425}
%\authornote{Equal contribution}
\affiliation{
\institution{Berlin, Germany, Spotify}
\city{}
\country{}
}
%\email{azing@spotify.com}

\author{Ekaterina Garmash\footnotemark[1]}
%\authornotemark[1]
%\orcid{0009-0008-3094-0892}
\affiliation{
\institution{London, UK, Spotify}
\city{}
\country{}
}
%\email{katyag@spotify.com}

\author{Gustavo Penha}
%\orcid{0000-0002-7373-0800}
\affiliation{
\institution{Amsterdam, Netherlands, Spotify}
\city{}
\country{}
}
%\email{gustavop@spotify.com}

\author{Kristen	Sheets}
%\orcid{0009-0005-8896-8597}
\affiliation{
\institution{San Francisco, US, Spotify}
\city{}
\country{}
}
%\email{ksheets@spotify.com}

\author{Martin Achenbach}
%\orcid{0009-0006-7391-4366}
\affiliation{
\institution{Berlin, Germany, Spotify}
\city{}
\country{}
}
%\email{machenbach@spotify.com}

\author{Oguz Semerci}
%\orcid{0009-0003-2889-9097}
\affiliation{
\institution{Boston, US, Spotify}
\city{}
\country{}
}
%\email{oguz@spotify.com}

\author{Remi Galvez}
% \orcid{0009-0003-6918-6923}
\affiliation{
\institution{New York, US, Spotify}
\city{}
\country{}
}
%\email{remigalvez@spotify.com}

\author{Marcus Tannenberg}
%\orcid{0000-0003-0077-4711}
\affiliation{
\institution{Gothenburg, Sweden, Spotify}
\city{}
\country{}
}
%\email{mtannenberg@spotify.com}

\author{Sahitya	Mantravadi}
%\orcid{0009-0002-3386-4246}
\affiliation{
\institution{New York, US, Spotify}
\city{}
\country{}
}
%\email{sahityam@spotify.com}

\author{Divya Narayanan}
%\orcid{0009-0001-3028-8186}
\affiliation{
\institution{New York, US, Spotify}
\city{}
\country{}
}
%\email{dnarayanan@spotify.com}

\author{Ofeliya	Kalaydzhyan}
%\orcid{0009-0004-9270-7018}
\affiliation{
\institution{Boston, US, Spotify}
\city{}
\country{}
}
%\email{ofeliyak@spotify.com}

\author{Douglas Cole}
%\orcid{0009-0004-1945-6857}
\affiliation{
\institution{Boston, US, Spotify}
\city{}
\country{}
}
%\email{dougcole@spotify.com}

\author{Ben	Carterette}
%\orcid{0000-0001-9538-047X}
\affiliation{
\institution{New York, US, Spotify}
\city{}
\country{}
}
%\email{benjaminc@spotify.com}

\author{Ann	Clifton}
%\orcid{0009-0003-4413-5257}
\affiliation{
\institution{New York, US, Spotify}
\city{}
\country{}
}
%\email{aclifton@spotify.com}

\author{Paul N. Bennett}
%\orcid{0009-0006-7852-9651}
\affiliation{
\institution{Boston, US, Spotify}
\city{}
\country{}
}
%\email{pbennett@spotify.com}

\author{Claudia	Hauff}
%\orcid{0000-0001-9879-6470}
\affiliation{
\institution{Delft, Netherlands, Spotify}
\city{}
\country{}
}
%\email{claudiah@spotify.com}

\author{Mounia	Lalmas}
%\orcid{0000-0002-3531-3096}
\affiliation{
\institution{London, UK, Spotify}
\city{}
\country{}
}
%\email{mounial@spotify.com}

%\renewcommand{\shortauthors}{Ghazimatin and Garmash, et al.}
\renewcommand{\shortauthors}{Azin Ghazimatin et al.}

\begin{abstract}
Listeners of long-form talk-audio content, such as podcast episodes, often find it challenging to 
understand the overall structure and locate relevant sections. 
A practical solution is to divide episodes into chapters—semantically coherent segments labeled with 
titles and timestamps. Since most episodes on our platform at Spotify currently lack creator-provided chapters, 
automating the creation of chapters is essential.  
Scaling the chapterization of podcast episodes presents unique challenges. First, episodes tend to 
be less structured than written texts, featuring spontaneous discussions with nuanced transitions. 
Second, the transcripts are usually lengthy, averaging about $16,$$000$ tokens, which necessitates 
efficient processing that can preserve context. To address these challenges, we introduce \model, a fine-tuned encoder-decoder transformer to segment conversational data. The model simultaneously generates chapter transitions and titles for the input
transcript. To preserve context, each input text is augmented with global context, including the 
episode’s title, description, and previous chapter titles. In our intrinsic evaluation, \model achieved a $11\%$ 
improvement in ROUGE score over the strongest previous baseline. Additionally, we provide 
insights into the practical benefits of auto-generated chapters for listeners navigating episode 
content. Our findings indicate that auto-generated chapters serve as a useful tool for engaging 
with less popular podcasts. 
Finally, we present empirical evidence that using chapter titles can enhance the effectiveness 
of sparse retrieval in search tasks.

\end{abstract}

%Additionally, 
%we confine boundary selection to sentences in the middle of a
%chunk, ensuring presence of 
%adequate context for generating meaningful and informative
%titles. 
%Comparing \model against several recent baselines, we
%observe $42\%$ improvement in Rouge score over GPT4 as the 
%strongest baseline.
%showcase its competitive performance according to segmentation and title generation metrics 
%across datasets from various domains.  
%?????\textcolor{red}{TODO: Say something about deployment at scale}.
%In addition to our intrinsic evaluation, we recently deployed the model on our platform. We find that
%users use our auto-generated chapters for browsing through 
%the episodes' content.
%\textcolor{red}{
%We further provide insights on usefulness of auto-generated chapters  for listeners to browse through episode content in practice.} We also present empirical evidence demonstrating the 
%benefit of leveraging chapter titles for improving the 
%effectiveness of sparse retrieval in search
%as a downstream task.
%treating the generated chapters
%as summaries of episode transcripts, we demonstrate
%that utilizing them for augmenting 
%episode descriptions
%yields  ~$8\%$increase in R@$30$ for sparse retrieval
%on the Podcast TREC dataset 
%compared to the baseline.

%\end{abstract}
\maketitle
\renewcommand{\thefootnote}{\fnsymbol{footnote}}
%\footnote{\small *Equal contribution.}
%\footnotetext{\small *Equal contribution.}
\footnotetext[1]{Equal contribution.}
\footnotetext[2]{Corresponding author. Email: azing@spotify.com}
\renewcommand{\thefootnote}{\arabic{footnote}}
\section{Introduction}\label{sec:introduction}
%We define "chapterization" as the task of identifying semantically coherent non-overlapping segments in a document and associating each identified segment with a proper title describing the content. Previous studies use the terms structured summarization~\cite{inan2022structured} or smart chaptering~\cite{retkowski2024text}  for the same task. The usefulness of document chapterization has long been recognized as a way to provide users with a convenient and structured content overview and simplifies navigation across the  document~\cite{gribbons1992organization, chelba2008retrieval}. Chapterization has also been shown  to serve as a useful intermediate step for other tasks, such as information  retrieval~\cite{shtekh2018exploring} and summarization  of long documents~\cite{xiao2019extractive, liu2022end}.  With the current rise in volume and availability of spoken user-generated content, such as podcasts and user-created videos,  chapterization is becoming increasingly more necessary with its content compression and navigational benefits \cite{chelba2008retrieval, jones2021current}. A recent in-house study found that users single out usefulness of chapters in podcasts for bookmarking, sharing and listening session planning.
%\footnote{In our user research, participants commented that they would like a feature for cataloging segments or moments of podcasts that resonated with them; a way to locate part of an episode so that they could share it with a friend; easy way of going back to part of a podcast and to get to a resource that had been suggested.}
We define \textit{chapterization} as the task of dividing a document into semantically coherent, non-overlapping segments and assigning each segment an appropriate title that reflects its content. This process, also referred to as structured summarization~\cite{inan2022structured} or smart chaptering~\cite{retkowski2024text}, has been shown to provide users with a 
convenient and structured content overview and simplify navigation across a 
document~\cite{gribbons1992organization, chelba2008retrieval}. The value of chapterization has been acknowledged for its role in facilitating other tasks such as information retrieval~\cite{shtekh2018exploring} and the summarization of lengthy documents~\cite{chelba2008retrieval, jones2021current}. With the increasing volume and availability of spoken user-generated content, like podcasts and videos, the need for chapterization has grown, offering significant benefits in content compression and navigation~\cite{chelba2008retrieval, jones2021current}. 
%Recent internal research highlights its utility in enhancing user experiences with podcasts by supporting bookmarking, sharing, and planning listening sessions.

%Podcast and video chapterization (aka timestamps) can be provided by content creators themselves -- however, it is often not the case, and moreover, there is no unified format or protocol for chapter annotation. Specifically on our platform that serves audio podcasts, less than $\sim$1\% of the episodes are chapterized by their creators. We bridge this gap by automatizing the chapterization process with a large language model-powered system trained on available creator chapters.

Podcast and video chapterization
%, often referred to as \textit{timestamps}, 
can ideally be provided by content creators themselves since there is no standardized format or protocol 
for chapter annotations. This, however, is frequently not the case; 
%and there is no standardized format or protocol for chapter annotations. On our platform that hosts audio podcasts, 
% less than  1\% of podcast episodes are chapterized by their creators. 
on our platform that hosts audio podcasts, the vast majority of episodes do not have creator-provided chapters.
We bridge this gap by automating chapterization using a large language model-powered system trained on available creator chapters.

\begin{figure}[t]
    \centering
    \includegraphics[width=0.9\columnwidth]{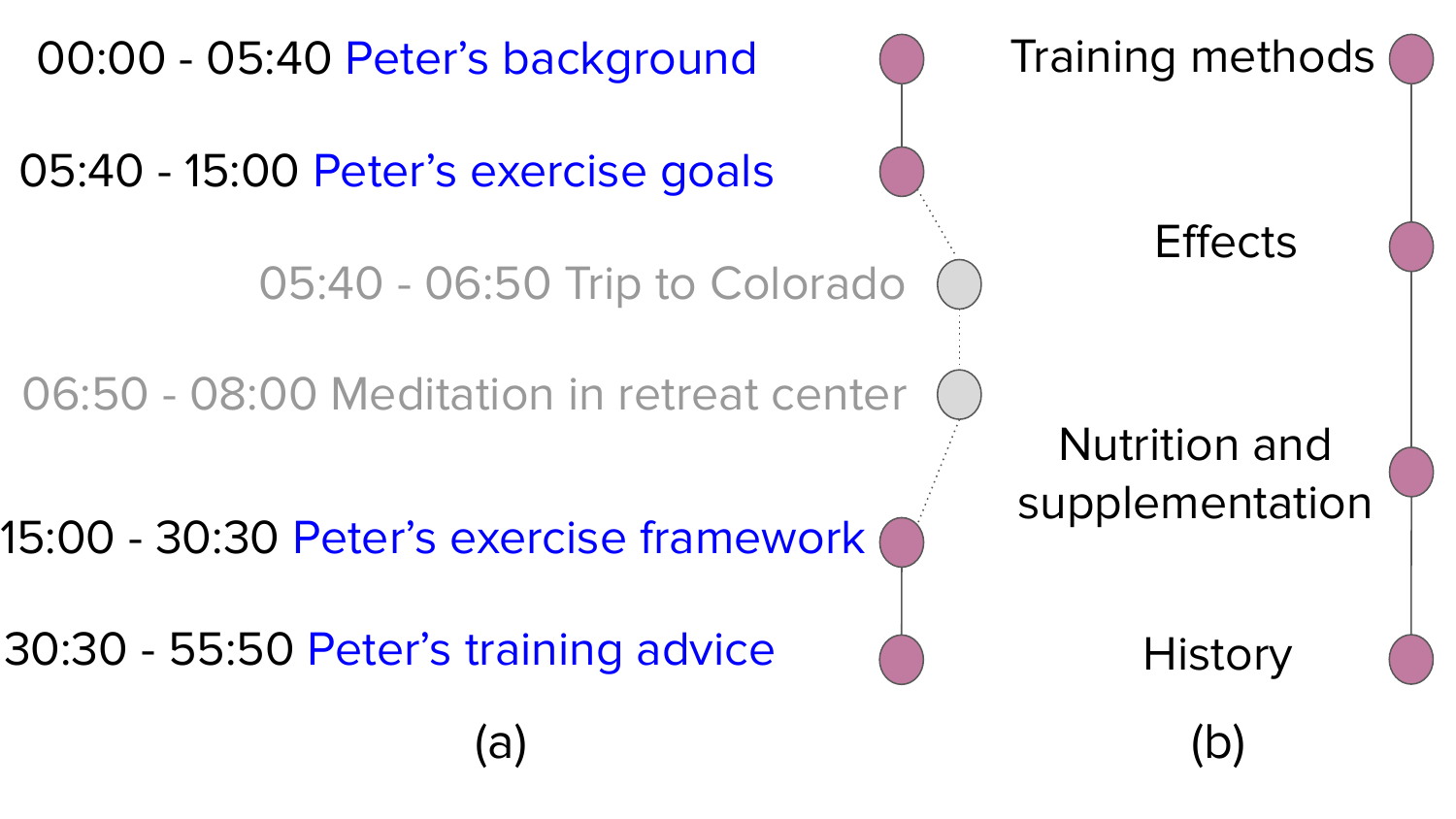}
    \caption{
    %Chapters for (a) a podcast episode about training tips 
    %vs. (b) a structured Wikipedia article about training and exercise. 
    %The podcast episode features (i) short tangential discussions (gray circles) within a broader topic, 
    %(ii) shared context (Peter’s training experience) between chapter titles, 
    %and (iii) a consistent style in titles. In contrast, Wikipedia chapters focus on the main topic with short titles that do not capture the global context.
    Chapters (purple circles) for (a) an episode about training tips 
    vs. (b) a structured Wikipedia article about training. 
    The episode chapters have short tangential discussions (gray circles), 
    shared context (Peter’s experience), 
    and a consistent title style. In contrast, Wikipedia chapters focus on the main topic with short titles that lack global context.
    }
    \Description{Comparison of the chapters of a podcast episode against those of a Wikipedia article.}
    \vspace{-0.3cm}
    \label{fig:intro}
\end{figure}

Most of previous research has concentrated on chapterizing structured written texts, such as Wikipedia articles, news, and journals~\cite{somasundaran2020two, lukasik2020text, liu2022end, 
yu2023improving}. There are however a few studies that focus on spoken discourse~\citep{zhong2021qmsum,inan2022structured, lin2023multi, retkowski2024text,yang2024vidchapters}. Yet, chapterizing spoken language documents, particularly podcast episodes, presents unique challenges compared to segmenting short, structured texts. Spoken discourse is usually more fluid, topically diverse, and less structured, and often features frequent digressions due to its interactive, real-time, and informal nature~\cite{jucker1992conversation, ghosh2022topic, retkowski2024text}.

Another challenge is the considerable length of podcast episodes, whether measured by time or word count when transcribed. This not only increases computational costs but also poses a modeling challenge; many podcasts contain long-range semantic dependencies that need to be captured by  chapterization. For instance, Figure~\ref{fig:intro}(a) shows a podcast episode where the discussion diverges into a tangent about traveling before returning to the main topic of exercising. Such tangents are typical of informal conversational podcasts. To predict a chapterization like the one in Figure~\ref{fig:intro}(a), a model must track the overarching context and theme. ``Knowing'' that the main topic is physical exercise helps the model distinguish segments about different aspects of this topic. Additionally, tracking predicted chapters throughout the episode helps the model generate consistent titles (in this example, 
focused on the guest named ``Peter'').
Chapterizing a Wikipedia article  as illustrated in Figure~\ref{fig:intro}(b), 
however, does not face these challenges since it is shorter and more structured.

Lately, there has been a growing focus on chapterizing conversational datasets.
In~\citep{inan2022structured}, segmentation and title assignment are modeled jointly, 
enhancing the predictive capabilities of both tasks. 
This model leverages LongT5~\cite{guo2021longt5} 
as the pre-trained 
sequence-to-sequence large language model (LLM). However, the context size of 
LongT5 is $16$k which is not sufficient for processing   
podcast transcripts with $16$k tokens on average. 
%This approach leverages a pre-trained sequence-to-sequence large language model (LLM). Its 
%ability to operate on a global context depends on the backbone LLM; \citep{inan2022structured} 
%uses LongT5, which has a relatively large context size of 16K tokens. 
In~\citet{retkowski2024text}, a two-stage chapterization model is used to first segment and then generate titles for the identified segments. This model uses longer context by incorporating previous chapter titles as left context summaries to generate chapter titles.
The model's two-stage design, however, inhibits information sharing between the 
two tasks. 

We can address the challenge of long inputs and long-distance dependencies in podcasts in several ways. First, a sufficiently large and powerful backbone LLM \textit{may} provide a large enough context window to process an entire episode’s transcript and produce accurate chapters. However, using a large LLM incurs significant computational and financial costs and may not fully capture all long-distance dependencies. 
To efficiently address these challenges, we propose \model, a chapterization model that
builds on the strengths of existing models, 
particularly~\citep{inan2022structured}, 
and extends them by dedicating a small portion of input text to explicit global context encoded as text: 
specifically, podcast episode metadata that reflects the overall content of the episode
and previously generated chapter titles. 
This allows a reasonably-sized\footnote{With less than a billion parameters.} LLM 
to handle long and unstructured content effectively, without solely relying on the LLM's power. Following \citep{inan2022structured}, we use LongT5 encoder-decoder model, which offers a compromise between efficiency and model power.
%Azin removed%Our approach therefore extends the core model~\citep{inan2022structured} with explicit global context encoded as text 
%Azin removed%input: specifically, podcast episode metadata and previously generated chapter titles. This allows a reasonably-sized\footnote{With less than a billion parameters} LLM to handle long and unstructured content effectively, without solely relying on the LLM's power. \textcolor{red}{Maybe say a bit more on the tech side?} 

We validate our proposed approach using two public non-podcast datasets and one internal 
podcast dataset. Our findings indicate that using global context as part of the input text
enhances the quality of chapter titles, particularly for longer documents 
in conversational datasets. 
We recently deployed \model 
%with FlanT5~\cite{chung2024scaling} 
on our platform. Usage statistics indicate that podcast listeners find 
the auto-generated chapters helpful for browsing through episodes, 
particularly in lesser-known podcasts.
Finally, we assess the utility of our 
generated chapter titles in a retrieval downstream task using the TREC 
Podcast Track dataset~\cite{jones2021trec}. Adding these titles to episode descriptions significantly enhances sparse retrieval effectiveness compared to an extractive summarization baseline.

We summarize our contributions as follows:
\squishlist
\item introduction of a new model, \model,  which effectively extends~\cite{inan2022structured} to address the challenges of podcast chapterization;
\item extensive intrinsic and extrinsic evaluations demonstrating the effectiveness and utility of the proposed approach;
\item deployment of the model in a user-facing production system and preliminary analysis of usage patterns for podcast chapters.
\squishend

\section{Related Work}\label{sec:related_work}
% \subsection{Unsupervised segmentation}\label{subsec:related_work_unsupervised}

We review related work, which has guided us in the various decisions we made to develop and deploy \model.
%\paragraph{Unsupervised text segmentation} These approaches  compute some type of a cohesion score or mutual  information~\cite{wagner2022topical} between  consecutive (blocks of) sentences utilizing TF-IDF  (or its variations)~\cite{hearst1997text, choi2000advances},  LDA topics~\cite{riedl2012topictiling}, probabilistic language  models~\cite{eisenstein2008bayesian, mota2019beamseg},  word2vec embeddings~\cite{alemi2015text}, or transformer based  embeddings~\cite{xing2021improving, solbiati2021unsupervised, gao2023unsupervised}.  The coherence scores are then plotted against the sentences, and  the valleys are considered as boundaries. When similarities are modeled as  edge weights in the semantic relatedness graph of the document, maximal cliques are treated as semantically coherent  segments~\cite{glavavs2016unsupervised}. 

\textbf{Text segmentation.} Early approaches for text segmentation (aka boundary detection) 
were unsupervised 
due to lack of sufficient supervised data. These approaches involve computing a cohesion score 
or mutual information~\cite{wagner2022topical} between consecutive blocks of sentences. 
This can be achieved using TF-IDF (or its variations)~\cite{hearst1997text, choi2000advances}, 
LDA topics~\cite{riedl2012topictiling}, probabilistic language models, 
word2vec embeddings~\cite{alemi2015text}, or 
transformer-based embeddings~\cite{eisenstein2008bayesian, mota2019beamseg}. 
The coherence scores are then plotted against the sentences, 
with the valleys considered as boundaries. 
When similarities are modeled as edge weights 
in the semantic relatedness graph of the document, 
maximal cliques are treated as semantically coherent 
segments~\cite{glavavs2016unsupervised}.
%\textcolor{red}{the link to next missing -- is it that these technqiues are less used as there is a move to supervided?}
%\textcolor{red}{we need to link back to the paper}
%\paragraph{Supervised text segmentation} Supervised approaches for text segmentation 
%typically train a boundary classifier on a sequence of input  sentences~\cite{tepper2012statistical, koshorek2018text, somasundaran2020two, zhang2021sequence, cho2022toward, xia2022dialogue},  or pairs of left and right context blocks~\cite{lukasik2020text, lee2023topic, vijjini2023curricular}.  To represent the input, these approaches employ statistical features~\cite{galley2003discourse, tepper2012statistical},  neural networks~\cite{sehikh2017topic,  wang2017learning, koshorek2018text, li2018segbot, arnold2019sector, xia2022dialogue, ding2022gits, zhang2019outline} or  transformers~\cite{somasundaran2020two, lukasik2020text, zhang2021sequence, lo2021transformer, li2022human, lee2023topic, xia2023sequence, lin2023multi, bai2023segformer}. 

%\paragraph{Supervised text segmentation}
The availability of large labeled data led 
to the increased use of supervised methods for 
addressing unique segmentation nuances across different domains.
These approaches generally involve training a boundary classifier on a 
sequence of input sentences~\cite{tepper2012statistical, koshorek2018text, somasundaran2020two, 
zhang2021sequence, xia2022dialogue, cho2022toward} or on pairs of left and right context 
blocks~\cite{lukasik2020text, vijjini2023curricular}. To represent the input, these 
methods utilize various techniques, including statistical features~\cite{galley2003discourse, 
tepper2012statistical}, neural networks~\cite{sehikh2017topic,  wang2017learning, koshorek2018text, 
li2018segbot, arnold2019sector, xia2022dialogue, ding2022gits}, or 
transformers~\cite{somasundaran2020two, lukasik2020text, zhang2021sequence, lo2021transformer, 
li2022human, lin2023multi, bai2023segformer}. 

Previous work on text segmentation primarily focuses on detecting 
segment boundaries without addressing title assignment which is 
necessary for podcast chapterization.
%These \textcolor{red}{both unsupervised and supervised?}methods only provide boundaries 
%without generating titles, which are essential for podcast chapterization. 
Next, we review 
previous studies that address both segmentation and title assignment.

%\textcolor{red}{we 
%need to link back to the paper}
\textbf{Joint segmentation and title assignment.} 
Prior studies suggest that jointly modeling the segmentation 
task and title assignment/generation offers mutual benefits for 
both tasks~\cite{inan2022structured}. In scenarios where the set of 
topics is considered closed, it is common practice to feed the learned representations 
into a multi-class classifier for 
title 
assignment~\cite{tepper2012statistical, arnold2019sector, lo2021transformer, gong2022tipster, lee2023topic, liu2023joint}. 
However, if the set of titles is open, a generative approach is employed~\cite{zhang2019outline, 
inan2022structured, liu2022end, xia2023sequence, lin2023multi}. 
%Given the state-of-the-art performance of previous studies on both conversational and written 
%texts~\cite{inan2022structured}, we leverage a generative model to jointly learn 
%segmentation and title generation. \textcolor{red}{not so clear how we contextualise the paper here still}
Given the diversity of podcast episode titles, we also adopt a 
generative approach similar to~\cite{inan2022structured}.

There is also a substantial body of literature on multi-modal segmentation~\cite{zhang2021sequence, ghinassi2023multimodal, yang2024vidchapters, xing2024multi}. However, since our model is uni-modal, we do not cover this topic  in this paper.

\textbf{Capturing long-range dependency.}
Chapters in a document can be seen as structured summaries of content~\cite{inan2022structured}, similar to a  summarization task. Therefore, chapterization is expected to benefit from capturing long-range context. Most recent studies aimed at making transformers more efficient for processing long texts focus on sparsifying or approximating the attention mechanism~\cite{beltagy2020longformer, wang2020linformer, kitaev2020reformer, choromanski2020rethinking, guo2022longt5, roy2021efficient, bertsch2024unlimiformer}. Another method for capturing long context is to hierarchically or incrementally merge the output of input chunks to facilitate information flow between them~\cite{chang2023booookscore}. However, this approach is slow and computationally expensive.
\citet{ge2023context} recently introduced in-context auto-encoders to compress long context into a few tokens, which can be passed as additional input to an LLM with a limited context window. Learning these tokens, however, requires extensive pre-training.
In our work, we use LongT5, which employs attention sparsification using global transient attention to efficiently capture longer context. Additionally, we augment the input chunks 
with document metadata to preserve context beyond the typical context size limit of most transformers.
\section{Method}\label{sec:model}
\begin{figure}[t]
\centering
    \centering
    \includegraphics[width=0.90\columnwidth]{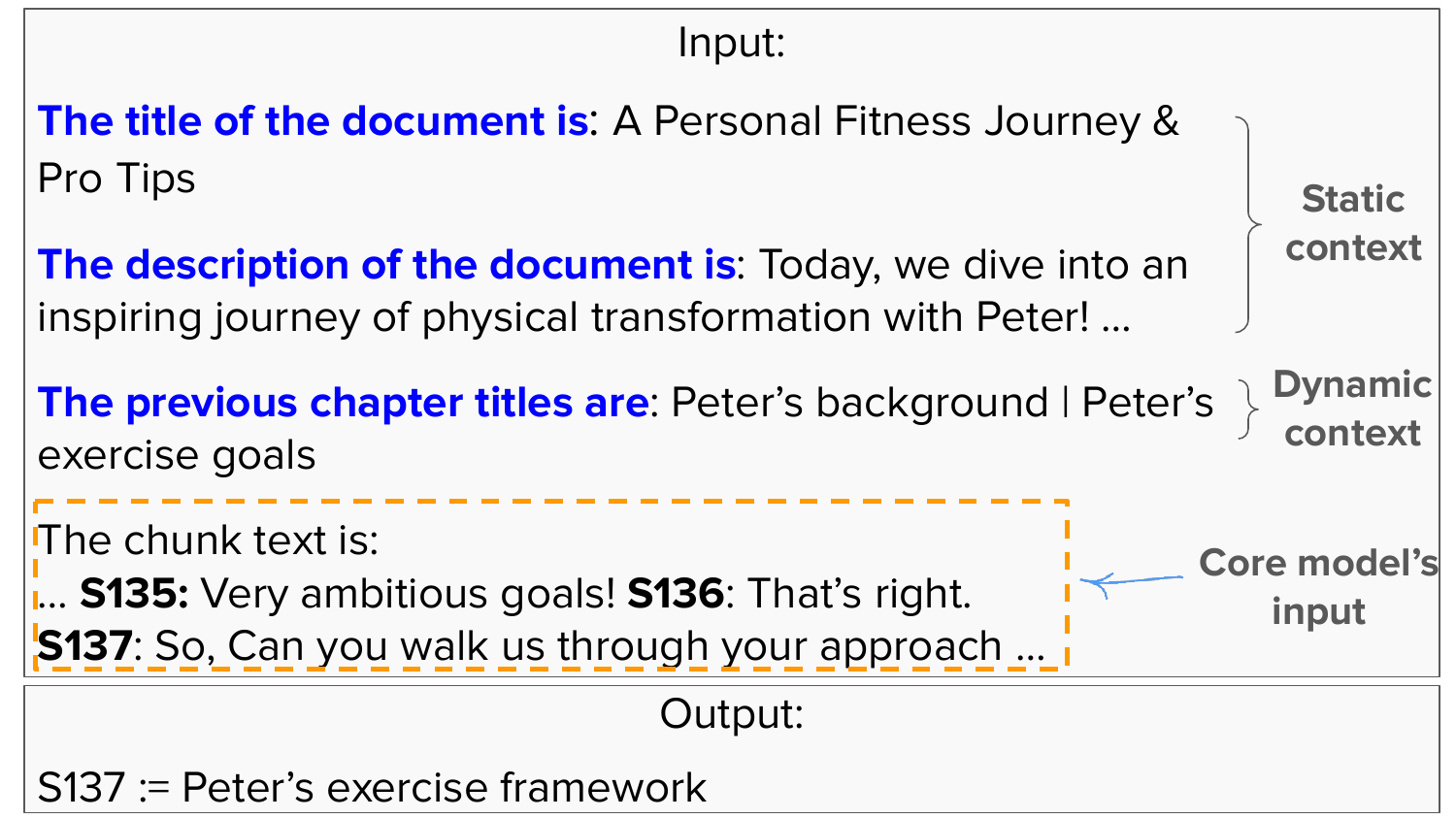}
    \caption{The input and output formatting of the chapterization model. The dotted box is the input to the core model.}
    \Description{The input and output formatting of the chapterization model. The dotted box is the input to the core model.}
    \vspace{-0.4cm}
    \label{fig:input-output}
\end{figure}

%Chapterization is a multifaceted task linked to many established NLP areas such as text segmentation, classification, and summarization, with various approaches emphasizing different aspects, as discussed in Section~\ref{sec:related_work}. 
Our work builds on the method by~\citet{inan2022structured}, modeling chapterization as simultaneous segmentation and title generation in a sequence-to-sequence fashion. The input is the text to be chapterized, and the output is a textual specification of chapter boundaries and titles. This approach uses a pre-trained LLM fine-tuned on supervised data, leveraging its vast linguistic and real-world knowledge. As a text-based model, it effectively integrates segmentation and title prediction, while also incorporating diverse contextual information to enhance the accuracy of the prediction, which is the main contribution of this paper.

%We refer to \citet{inan2022structured} as the core model: we explain its internals and how we apply it to podcast domain in Sec.~\ref{sec:core_model}. Our contribution consists in incorporating additional contextual cues in the core model, with the aim of improving generalization on long-input data and  alleviating the disadvantages of local nature of  chapterization inference. Specifically, we explore:

We refer to the model by~\citet{inan2022structured} as our core model, which we detail and explain its application to the podcast domain in Section~\ref{sec:core_model}. Our contribution involves incorporating additional contextual cues into this core model to improve generalization on long-input data and mitigate the limitations of the local nature of chapterization inference. Specifically, we explore: 
\squishlist
%\item[(1)] \textbf{static context} (Sec.~\ref{sec:doc_context}): metadata outlining the \textit{overall} content of the input document. It can be useful in cases where the full document cannot be accessed by the model in one go. Concrete instantiations depend on the domain and dataset, we specify them in Sec.~\ref{sec:exper_setup}.
\item[(1)] \textbf{Static context} (Section~\ref{sec:doc_context}): Metadata outlining the \textit{overall} content of the document. This is useful when the model cannot access the entire document at once. Specific implementations depend on the domain and dataset, detailed in Section~\ref{sec:exper_setup}.
%the overall summary of the input document. This  information is useful for cases where the full document cannot be accessed by the model in one go, which is typically the case for podcasts and most SoTA LLMs. We discuss the various proxies to document summary we can use.
%\item[(2)] \textbf{dynamic context} (Sec.~\ref{sec:dynamic_context}): the intermediate state of the left-to-right chapterization process of the given document. This type of information provides access to the chapterization decisions made at earlier steps and thus enables the model to guide the choice of the following chapters.
\item[(2)] \textbf{Dynamic context} (Section~\ref{sec:dynamic_context}): The intermediate state of the left-to-right chapterization process. This information provides access to earlier chapterization decisions, guiding the selection of subsequent chapters.
%``reason'' 
%better in terms of full search paths.
\squishend
%Figure~\ref{fig:input-output} illustrates the input formatting with both types of context augmentation.

\subsection{Core model}
\label{sec:core_model}
%Our core model is roughly equivalent to $\mathit{l}^{\text{GEN}}_{\text{seg + label}}$ from \citet{inan2022structured}, which was experimentally shown to be the best performing variant of their model. The model has a encoder-decoder architecture with an underlying Transformer LLM. Any existing LLM conforming to the seq2seq API can be  used in this approach, but in our experiments we use the LongT5 pre-trained LLM \citep{guo2021longt5}, following \citet{inan2022structured}. 

Our core model is based on the segmentation and labeling framework \textbf{Gen (seg+label)} from~\citet{inan2022structured}, which has been demonstrated experimentally to be the best-performing variant of their model. This model employs an encoder-decoder architecture with an underlying Transformer LLM. While any existing LLM adhering to the seq2seq API can be used, our experiments specifically use the LongT5 pre-trained LLM~ \citep{guo2021longt5}, in line with~\citet{inan2022structured}.

%The exact input-output formatting of the core model is shown in Fig.~\ref{fig:input-output}. We augment the raw input text by adding index numbers before each sentence, so that the decoder can predict the start of a chapter by referring to one of those indices. The output sequence is a chronologically ordered concatenation of strings of the form:

The input-output formatting for our core model is illustrated in Figure~\ref{fig:input-output}. We augment the raw input text by adding index numbers before each sentence. This allows the decoder to predict the start of a chapter by referencing one of these indices. The output sequence is a chronologically ordered concatenation of strings formatted as:
\begin{itemize}
\item[] \verb|${first_sentence_index} := ${title}|
\end{itemize}
%
 %Since input text can be arbitrarily long (especially for such media as podcasts, see Table~\ref{tab:data_stat}) and modern open source LLM input capacity is limited\footnote{LongT5 context size is 16,384, see \url{https://huggingface.co/docs/transformers/en/model_doc/longt5}.}, a necessary first step of our approach is  chunking of the input text that produces spans of size directly consumable by a LLM. Each training datapoint is therefore a chunk of input text and the output string with chapter boundaries and titles corresponding to that chunk. If the given chunk does not contain chapter boundaries, the output string is \texttt{"No chapter boundaries were found."}. We do chunking with a sliding non-overlapping window of size smaller than the given LLM's input capacity. 
 %depending on which results in the majority of datapoints (chunks) having a "No chapter boundaries were found"\footnote{Average episode length in words is TODO, average number of chapters per episode is TODO, LLM encoder input size varies from 300 to 1500 -- TODO double-check numbers.}, which may have negative effect on the model due to overfitting. We experimented with various resampling techniques, such as negative class (No boundaries) class downsampling, to control the ``eagerness'' of model to find a chapter boundary in a given chunk at test time.
%
 with character ``$|$'' as the separator. Given that input texts can be arbitrarily long--particularly in media such as podcasts (see Table~\ref{tab:data_stat})--and 
 %\textcolor{red}{at the times of writing} 
 open-source LLMs have limited input capacities,\footnote{LongT5 context size is 16,384, see \url{https://huggingface.co/docs/transformers/en/model_doc/longt5}.} the initial step in our approach involves chunking the input text into segments that can be processed by an LLM. Each training datapoint consists of a chunk of input text and the corresponding output string, which includes chapter boundaries and titles relevant to that chunk. If a chunk does not contain any chapter boundaries, the output string is \texttt{"No chapter boundaries were found."} The chunking process uses a sliding non-overlapping window with a size smaller than the LLM’s input capacity.
 
 This necessity for chunking the input can result in predictions that are locally informed and are not based on the broader context about the entire input text or about the chapterization predictions made in the preceding chunks. 
 Given the considerable average length of podcasts and the frequent presence of long-distance dependencies, such locality may result in suboptimal chapter quality.
 To address this limitation, we propose incorporating global context using methods described below.
 
%The necessity to chunk the input is bound to result in a locally informed model, and it may not do well in cases where wider context -- either about the content of the raw input text, or the predictions already made --  is necessary to make the right prediction. We propose two methods to alleviate this, below.
 
\subsection{Adding static context}
\label{sec:doc_context}
%In a chunked input processing (explained above), the model does not have access to what content comes before or after the given chunk. This may result in predicted chapters not being specific enough to differentiate content outside of the chunk where the given prediction is made. Or, vice versa,  the chapter boundaries and titles may single out details prominent in the given chunk but irrelevant to the general discussion or narration. 

When processing chunked input (explained above), the model lacks access to content before or after a given chunk. This can result in predicted chapters that are either not  specific enough to distinguish content outside the chunk or too focused on details specific to the chunk but irrelevant to the overall discussion. 

%We propose to include a piece of data that outlines the overall content or structure of the whole document, thus providing general context about the document. We refer to it as \textit{static context}, because it is given prior to chapterization and it does not change. The exact content and structure of such metadata may vary from dataset to dataset, and we specify our exact choices in Sec.~\ref{sec:exper_setup}. In Fig.~\ref{fig:input-output} we give an example where as static context we use the title and the description of a podcast episode.

To address this, we propose including metadata that outlines the document's overall content, providing a general context. We call this \textit{static context}, as it is provided prior to chapterization and remains unchanged. The specific content and structure of this metadata varies by dataset, detailed in Section~\ref{sec:exper_setup}. Figure~\ref{fig:input-output} shows an example with the title and description of an episode as static context.

\begin{comment}
We propose to include a short summary of the whole document to provide the model with such general context about the document. We consider the following sources and proxies to document summaries:
\squishlist
\item General end-to-end summary: it rarely exists in the wild, but some curated datasets are annotated with short per-document summaries (among the datasets used in the experiments, QMSum \citep{zhong2021qmsum} comes with human-written summaries).
\item Document title: titles can be seen as highly ``compressed'' summaries and are typically available for documents of various domains.
\item Episode description: podcast episodes usually come with a creator-provided description. Note that episode descriptions are usually different from general summaries in that they only provide a high-level outline of the topics of the discussion with the intention of enticing the user to listen to the episode.
\squishend

In Sec.~\ref{sec:data} we specify which summary proxies are available for which datasets in our experiments. 
\end{comment}

\subsection{Adding dynamic  context}
\label{sec:dynamic_context}
%Another disadvantage of local chunked processing is that the model is not aware of the chapterization decisions it has made at earlier steps for the given input document. The consequence is that for each local prediction step the boundaries and titles may turn out to be inconsistent with previous decisions. Examples of inconsistencies include repetitive titles, different levels of chapter granularity, varying linguistic styles of titles. 
%We analyze 
% such ``inconsistency errors'' 
%the impact of dynamic context on the overall coherence of chapter titles in Sec.~\ref{sec:experiments}.
%(pre- and post- model context modification) in Sec.~\ref{sec:experiments}.

Another disadvantage of local chunked processing is the model's lack of awareness of prior chapterization decisions for a given input document. As a result, each local prediction step may produce boundaries and titles that are inconsistent with previous decisions. This can lead to issues such as repetitive titles, different levels of chapter granularity, and varying linguistic styles in titles.

%As dynamic information about the state of the chapterization process, we add to the input text the sequence of titles already predicted for the earlier portion of the document. 
%We experiment different memory lengths: from fixed $k$ previous titles to all previously predicted titles. 
%Figure~\ref{fig:input-output} illustrates how previously predicted titles are incorporated into the input text.

To provide dynamic information about the state of the chapterization process, we add the sequence of titles already predicted for the earlier portions of the document to the input text. Figure~\ref{fig:input-output} shows how  previously predicted titles are added to the input text.

\section{Experimental setup}
\label{sec:exper_setup}
\begin{table*}[ht]
\centering
%\small
%\begin{tabular}{l l l l l l}
\caption{Statistics for datasets used in the experiments. The terms ``doc'' and ``desc'' denote  
document and description, respectively.\label{tab:data_stat}}
\footnotesize
\begin{tabular}{l P{1cm} P{2cm} P{2cm} P{2cm} P{2cm} P{4cm}}%{lP{1.5cm}P{1.5cm}P{1.5cm}P{1.5cm}P{1.5cm}P{1.5cm}}
%\begin{tabular}{lp{1.5cm}p{0.5cm}p{1.5cm}p{1.5cm}p{1.5cm}|p{0.5cm}p{0.5cm}p{1.5cm}p{1.5cm}p{1.5cm}}
\toprule
 %\textbf{Dataset} (train size) & No. chapters p/episode &\multicolumn{2}{l}{\textbf{Segments}} & \multicolumn{3}{l}{\textbf{Titles}} \\
 %& & \multirow{1.5}{*}{\parbox{2cm}{Length \\(in sentences)}}  & \multirow{2}{*}{\parbox{3cm}{Avg. no. tokens}} & Length (in words) & \multirow{2}{*}{\parbox{3cm}{\textbf{Example titles}}} \\
\textbf{Dataset} (size) & \textbf{Avg. no. chapters} & \textbf{Avg. segment length} (sentences) 
& \textbf{Avg. title length} (words) & \textbf{Avg. no. tokens (doc title desc)} & \textbf{Avg. no. words (doc title desc)}& \textbf{Example title}
\\
\toprule
Podcast (10,804) & 11.3 $\pm$ 7.5 & 80.9 $\pm$ 83.7  & 6.2 $\pm$ 5.7 & 16,098 \phantom{0} 20 \phantom{0}130 & 11,845 
 \phantom{0} 11 \phantom{0}102&\footnotesize How your smile affects others\\
%\rowcolor{verylightgray} 
Wikisection (23,129) & 5.2 $\pm$ 3.8 & 11.0 $\pm$ 13.7 & 1.0 $\pm$ 0.0 &  1,603 \phantom{00} 6 \phantom{0} 134 & 1,068 \phantom{00} 2 \phantom{00} 85&\footnotesize city.geography, disease.genetics \\
QMSum  (232)  & 4.7 $\pm$ 1.9 & 163.3 $\pm$ 156.9 & 6.4 $\pm$ 3.6 & 13,391 \phantom{0} 0 \phantom{0} 130 & 8,467 \phantom{00} 0 \phantom{0} 104 & \footnotesize Design and availability of actual components \\
\bottomrule
\end{tabular}
\vspace{-0.3cm}
\end{table*}

%We describe the datasets, baselines and metrics used to validate \model. We also provide some implementation details. 
\subsection{Datasets}
\label{sec:data}
We downsampled our \textbf{podcast dataset} from a proprietary internal catalog, using only English episodes that were chapterized by their creators. 
Our final dataset contains $10.8$k episodes, 
uniformly sampled with several filters. 
Chapters in these episodes range from $30$ seconds 
to $30$ minutes, and titles are shorter than 
$15$ words.  We randomly split the resulting dataset 
into train/validation/test partitions of $8$k/$1$k/$1$k episodes. 
For each episode, we use both title and description as the static context
since $96\%$ of episodes in our catalog have descriptions, 
with $57\%$ of those longer than $20$ words. The majority ($91\%$) 
of episodes in our dataset are conversational, 
featuring multi-speaker discussions.

To gauge \model's effectiveness across different domains, we use two other publicly available English datasets. \textbf{WikiSection}~\citep{arnold2019sector} is a Wikipedia-based dataset limited to two categories, \textit{en\_disease} and \textit{en\_city}, with normalized section titles for discriminative title prediction. Examples of segment titles are in Table~\ref{tab:data_stat}. We use only the English documents and use the title and abstract of each document as the static context. The second dataset, \textbf{QMSum}~\citep{zhong2021qmsum}, is a collection of meeting transcripts annotated with topic segments and labels. 
While it is closer to podcasts as it involves conversational data, it is a low-resource dataset with only $232$ data points. We use the user-generated meeting summaries as the static context.

Table~\ref{tab:data_stat} presents descriptive statistics for the three datasets. %, highlighting domain differences. 
Compared to Wikisection, the podcast and QMSum datasets feature longer documents and chapters, with more descriptive chapter titles. The podcast dataset shows significant variability in the number of chapters per episode and title length, indicating greater diversity.

\subsection{Baselines}\label{subsec:baselines}
We compare \model against the following baselines:
%that use only text input: 
\squishlist
    % \item \textbf{TSSP+CSSL}
   % \item[] \textbf{CATS}~\cite{somasundaran2020two}: A multi-task learning model that combines  the boundary classification task with coherent sequence detection, whereby the model learns to differentiate correct sequences of sentences from the corrupt ones.  This choice of this model as a baseline is motivated by the recent state-of-the-art performance  of hierarchical encoders for segmenting video  transcripts~\cite{retkowski2024text}.
     \item[] \textbf{CATS}~\cite{somasundaran2020two}:  A multi-task learning model that combines boundary classification with coherent sequence detection 
     %The model is trained to 
     that differentiate correct sequences of sentences from 
     the corrupt ones. This model is chosen due to the recent state-of-the-art performance of hierarchical encoders for segmenting video transcripts~\cite{retkowski2024text}.
    %To learn sentence representations, CATS employs stacks of transformers.
    %Note that this model does 
    %not generate or detect topic labels, and thus we report only the boundary metrics 
    %for it.
  %  \item[] \textbf{Gen (seg + label)}~\cite{inan2022structured}: A single-stage   seq2seq model   that uses LongT5 to jointly generate     chapter titles and boundaries    (structured summarization). This model is similar to \model's core model.
    \item[] \textbf{Gen (seg + label)}~\cite{inan2022structured}: A single-stage seq2seq model that uses LongT5 to jointly generate chapter titles and boundaries (structured summarization), similar to \model's core model.
    %In this model, the sentence ids are encoded using the first $n$ vocabulary 
    %of the tokenizer where $n$ denotes the largest index for a sentence. 
    %Given the large size of the transcript episodes, we implement chunking 
    %where we break the input text into 
    %chunks of
    %$8000$ words.
    %%the same 
    %%chunking mechanism as ours in this method.
  %  \item[] \textbf{GPT-4}~\cite{achiam2023gpt}: We perform zero-shot learning     with GPT-4 with extended context     of $128$k tokens     (gpt-4-0125-preview),   where we instruct the model to chapterize the entire    transcript    and return
    %Considering the sufficiently large context window of 
    %this model, we did not employ chunking and used the entire transcript 
    %in the prompt. We instructed the model to 
%    the chaper titles and 
%    the IDs of the starting sentences in JSON format for easy parsing.
 %   The experiment was run in the 
%    second week of May 2024.
    \item[] \textbf{GPT-4}~\cite{achiam2023gpt}: Zero-shot learning with GPT-4, using an extended context of 128k tokens (gpt-4-0125-preview). We instruct the model to chapterize the entire transcript and return the chapter titles and starting sentence IDs in JSON format for easy parsing. The experiment was conducted in the second week of May 2024.
    %Given the cost associated with 
    %querying this model, we compare its performance against our 
    %model only on a randomly selected subset of the podcast dataset.
\squishend

\subsection{Implementation details}
%As our backbone model we use LongT5 \citep{guo2021longt5} with \texttt{base} size ($\sim$$250$M parameters). We use the transient-global attention setting. We used train batch size of 
%$8$, constant learning rate of $5.0e$-$5$, and maximum of $8$ epochs. 
%and warm up period of $10$k steps. 
% early stopping grace period of 2. 
%We divide the input transcripts into non-overlapping chunks of $500$ words. Given the higher frequency of negative chunks (chunks without chapter transitions), we down-sample these during the training. We set the ratio of negative examples in our training data to $0.4$ which has been shown empirically to prevent over-chapterization at lower ratios and under-chapterization at higher values. The training and inference for offline evaluations were done on a Ray~\cite{moritz2018ray} cluster with a single node and a single GPU. One full training on podcast dataset took on average $\sim$$3$ days. On the same, inference of $1.1$k episodes with and without dynamic context takes on average $1$ hour and $3$ hours, respectively. Note that with dynamic context, it is still feasible to do batch inference since the $ith$ chunks of the episodes can be processed all together. 

We use LongT5~\citep{guo2021longt5} (base size, $\sim$$220$M 
parameters) with transient global 
attention, as our backbone model.  The training setup includes a batch size of $1$, a 
learning rate of $5.0e$-$5$ with scheduler type of linear, 
and a maximum of $4$ epochs. The same setting was used for 
other datasets with the exception of learning rate 
$1.0e$-$4$ for Wikisection. We 
use input chunks of up to $8000$ words,\footnote{A conservative choice to ensure there are no more 
tokens than $16$k.} 
with $7000$ words 
dedicated to the document text 
and up to $1000$ words to the metadata. In Gen(seg+label),
all the $8000$ words are used for document text.
On average, 
each transcript in the podcast dataset is broken into $1.75$ chunks.
%To address the higher frequency of negative chunks 
%(those without chapter transitions), we downsample them during 
%training. We set the ratio of negative examples in our training 
%data to 0.4, which has been empirically shown to prevent over-
%chapterization at lower ratios and under-chapterization at 
%higher values. 
%
Training and inference for offline evaluations were conducted on a Ray~\cite{moritz2018ray} cluster with a single node and a single GPU. Training on the podcast dataset took approximately $3$ days. 
%Training on the podcast dataset took approximately 3 days. 
Inference of $1.1$k episodes lasts an average of $1$ hour.

\subsection{Evaluation metrics}
%To the best of our knowledge, there are currently no metrics that evaluate chapterization holistically. 
%Given the compositional nature of the chapterization task, it is common to evaluate chapter boundaries and generated titles separately using their corresponding metrics~\citep{inan2022structured}. For segmentation evaluation, we use \textbf{WindowDiff}~\cite{pevzner2002critique} which is the average difference between  the number of boundaries in  predicted and reference values computed over  the spans of $k$ sentences. This metric is parametrized by $k$,  the size of the sliding window, which is usually set to half of the average segment  length (in sentences).  We estimate $k$ for each dataset using the train partition and report it in  the results table (Table~\ref{tab:com-table}). Lower values of this metric indicate more accurate segmentation.

%Given the compositional nature of the chapterization task, 
It is common to evaluate chapter boundaries and generated titles separately using their respective metrics~\citep{inan2022structured}. 
For segmentation evaluation, we use \textbf{WindowDiff}~\cite{pevzner2002critique}, which measures the average difference between the number of boundaries in predicted and reference values over spans of $k$ sentences. This metric is parametrized by $k$, the sliding window size, usually set to half the average segment length (in sentences). We estimate $k$ for each dataset using the training partition and report it in Table~\ref{tab:com-table}. Lower metric values indicate more accurate segmentation.

%For titles, one can use reference-based summarization metrics, such as \textbf{ROUGE} \citep{lin2004rouge} and BERTScore \citep{zhang2019bertscore}. Previous work such as~\cite{inan2022structured} use summarization metrics  at document level, i.e., they compute metrics on summaries created by concatenating  chapter titles in chronological order. This evaluation approach, however, hinders  individual  title assessment.
%To evaluate titles individually, we employ a heuristic alignment method between reference 
%and prediction chapters. For each chapter $c_i$
%in one set (reference or prediction), we find a chapter $c_j$ in 
%the other set with highest overlap. Then, we match the title of chapter $c_i$ 
%in one set
%with the title of chapter $c_j$ in the other set.
%with the title of a chapter in the other set 
%that has the highest segment overlap. 
%Note that this matching process is asymmetric-when a title from the reference set is matched to a title in the prediction set, the reverse match is not guaranteed. We use SBERT title representations  \citep{reimers2019sentence} \footnote{Note that we use SBERT and not BERT for title representation since we measure distances between whole chapterization lists where the atomic elements are titles -- unlike, for example, BERTScore, which was originally designed to measure similarity between sentences, where the atomic elements are words, in a machine translation and image captioning scenarios.} to apply soft-matching distance and define the metrics as:
% chapter titles are concatenated in chronological order to form a single ``summary''. 

For titles, reference-based summarization metrics like \textbf{ROUGE} \citep{lin2004rouge} and 
BERTScore~\citep{zhang2019bertscore} are commonly used. Previous work often computes these 
metrics on summaries created by concatenating chapter titles sequentially, which 
hinders individual title assessment. To evaluate titles individually, we employ a heuristic 
alignment method between reference and predicted chapters. For each chapter $c_i$ in one set 
(reference or prediction), we find the chapter $c_j$ in the 
other set with the highest overlap at sentence level, 
then match their titles. Note that this matching process is asymmetric, meaning a title matched from the reference to the prediction set does not guarantee a reverse match. We use SBERT title representations 
\citep{reimers2019sentence}\footnote{We use SBERT instead of BERT for title representation because we measure distances between entire chapter lists, where the atomic elements are titles. Unlike BERTScore, which measures similarity between sentences at the word level for tasks like machine translation and image captioning, SBERT is better suited for our purpose.} to apply soft-matching distance and define the metrics as:
\begin{equation}
\text{ROUGEL$_{F1,aligned}$} = \frac{\sum_{(t, t')\in \text{Matches}_{\text{all}}} 
\scriptstyle\text{ROUGEL$_{F1}$}(t, t')
}{|\text{Matches}_{\text{all}}|}
\end{equation}
\begin{equation}\label{eq:sbert_pred}
    \text{SBERT}_{\text{prec}} = \frac{\sum_{(t, t')\in \text{Matches}_{\text{pred}}} \scriptstyle\text{SBERT}(t, t')}{|\text{Matches}_{\text{pred}}|}
\end{equation}
\begin{equation}\label{eq:sbert_recall}
    \text{SBERT}_{\text{recall}} = \frac{\sum_{(t, t')\in \text{Matches}_{\text{ref}}} \scriptstyle\text{SBERT}(t, t')}{|\text{Matches}_{\text{ref}}|}
\end{equation}
where $Matches_{pred}$ is the set of title pairs $(t, t')$ where a predicted title $t'$ is matched with 
reference title $t$ with highest overlap. Similarly, $Matches_{ref}$ is a set of title pairs $(t, t')$
where reference title $t$ is matched with predicted title $t'$ with highest overlap. 
$Matches_{all}$ denotes the union of $Matches_{pred}$ and $Matches_{ref}$. For simplicity, 
we refer to ROUGEL$_{F1,aligned}$ as ROUGEL$_{F1}$ hereinafter.
%is the set and reference chapter titles for one episode, respectively. \textit{MaxOverlap} is a function that takes a title $t$ and maps the corresponding segment $s$ to a maximally overlapping segment in set $T$.
\textbf{SBERT$_{F1}$} is computed as the geometric mean of (\ref{eq:sbert_pred}) and (\ref{eq:sbert_recall}).

\subsection{Ethics Statement}\label{subsec:ethical_statement}
%Podcast creators are in the best position to chapterize their own content. Therefore, 
%we do not generate chapters for episodes whose creators have already provided chapters. 
%Besides, we do not display chapters for the content of the creators who opt out from the 
%auto-generated chapterization experience. We also inform users about the fact that chapters 
%are generated by AI using the following disclaimer: \textit{The chapters are auto-generated.}

%We display auto-generated chapters for episodes without creator chapters and inform users about the fact that chapters are generated by AI using the  following disclaimer: \textit{The chapters are auto-generated.}  In addition, we ensure compliance with terms and conditions of Spotify  for Podcasters and enable creators to opt out of  the experience at their discretion. 

We display auto-generated chapters for episodes that do not have creator-provided chapters. Users are informed that these chapters are generated by AI with the following disclaimer: \textit{The chapters are auto-generated.} Additionally, we ensure compliance with the terms and conditions of Spotify for Podcasters and allow creators to overwrite AI-generated chapters or opt-out of this feature at their discretion.
%To safeguard users from exposure to possibly harmful content generated by LLMs, we employ a safety mechanism where we remove sensitive or inappropriate titles before  being displayed to the users. In addition,  we also allow for the immediate manual removal of  harmful content if any is reported.
To protect users from potentially harmful AI-generated content, we employ a safety mechanism to remove sensitive or inappropriate titles before they are displayed. We also allow for the immediate manual removal of any reported harmful content.

\section{Offline Results}\label{sec:experiments}

We present the findings from our experiments, addressing four research questions. The results are detailed in Tables~\ref{tab:com-table},~\ref{tab:long-short} and~\ref{tab:anecdotal-examples}. 
%\textcolor{red}{to check} %\textcolor{red}{Call this section Offline Results?}
%We present our findings for the following four research questions:
\\
\begin{table*}[t]
\centering
\footnotesize
%\begin{tabular}{l l l l l l}
\caption{Comparison of \model against the baselines according to boundary and title 
metrics across three datasets: Podcast, Wikisection, and QMSum. The best metric values for each dataset are marked in bold. $^\ddagger$ denotes statistical significance
of \model over the strongest baseline. $\downarrow$ indicates that lower values are better. $k$ denotes the parameter value of \textit{WinDiff}. Notation 7000+1000 (chunk size) means that 7000 words are used for input document text and 1000 for static and dynamic context.}
\begin{tabular}{llP{2cm}P{2cm}P{1.5cm}P{1.5cm}P{1.5cm}}
% \begin{tabular}{lcccc}
%\begin{tabular}{lp{1.5cm}p{0.5cm}p{1.5cm}p{1.5cm}p{1.5cm}|p{0.5cm}p{0.5cm}p{1.5cm}p{1.5cm}p{1.5cm}}
\toprule
%& \multicolumn{5}{c|}{Podcast dataset}   &  \multicolumn{5}{c}{Another conversational dataset?}
% \\ \midrule
%& \multicolumn{2}{l}{Boundary metrics} & \multicolumn{4}{l}{Title metrics} \\
%\midrule
& Model & Chunk size (words)   & Static/Dynamic context & WinDiff $\downarrow$    & ROUGEL$_{F1}$$\uparrow$  & SBERT$_{F1}$$\uparrow$ \\
%&  pk    & WD   & sBERTs Precision & sBERTs Recall & sBERTs F1
 \hline
%\textbf{Podcast dataset} $k$=$45$ &  &  &  &    &  & \\
% TSSP+CSSL& 0.49 & 0.60  & - & -  & - & - \\
%\rowcolor{verylightgray} 
\multicolumn{7}{c}{Podcast dataset ($k=45$)} \\
\hline
(1)    & CATS~\cite{somasundaran2020two} & - & - & 0.505   & -  & - \\
(2) & GPT-4~\cite{achiam2023gpt} &  - & \cmark/\xmark & 0.448 & 0.134 & 0.315\\
(3) &  Gen (seg+label)~\cite{inan2022structured} & 8000 & \xmark/\xmark & \textbf{0.364} & 0.208 &  0.394 \\
%PODTILE & 4 &   4000 &  \cmark/\cmark & \textbf{0.389}$^\ddagger$ & \textbf{0.215}$^\ddagger$ &  \textbf{0.396}$^\ddagger$ \\
(4) & \model &   7000+1000 &  \cmark/\cmark & 0.365 & 0.231$^\ddagger$ &  0.414$^\ddagger$ \\
%\cdashline{3-8}
\cline{1-7}
% 0.43    0.34    0.14    0.32    0.33    0.32
% Ours (no expansion)    & 0.32 &  0.44  & 0.25 & 0.51   & 0.54 & 0.52 \\
% 0.18    0.44    0.32    0.25    0.51    0.54    0.52
%Ours (no desc)    &  &    & - & -   & - & - \\
%Ours (non-sequential)    &  &    & - & -   & - & - \\
%\multirow{3}{*}{PODTILE (Ablation)} & 5 &   4000 & \cmark/\xmark &  0.391 & 0.222 & 0.400 \\
(5) & \multirow{3}{*}{\model (Ablation)} &   7000+1000 & \cmark/\xmark &  0.368 & \textbf{0.235}$^\ddagger$ & \textbf{0.418}$^\ddagger$ \\
(6) &  &   7000+1000  & \xmark/\cmark & 0.368 & 0.209 & 0.392 \\
(7) &  &   7000 & \xmark/\xmark & 0.371 & 0.215$^\ddagger$ & 0.400$^\ddagger$ \\ 
%\hdashline
%(8) & \model (FlanT5-base) &   300 & \xmark/\xmark & 0.466 & 0.156 & 0.328 \\ 
%(9) &  \model (FlanT5-large) &   300 &  \xmark/\xmark & 0.472 & 0.185 & 0.364 \\ 
%\hdashline
%\model$\setminus$Static\&Dynamic (500)  &  0.31 &  0.43 & 0.18 & 0.36 &  0.37 & 0.36 \\ 
%\model$\setminus$Static (500)  &  0.33 &  0.47 & 0.17 & 0.34 &  0.37 & 0.35 \\ 
%\model$\setminus$Dynamic (500)  &  0.31 &  0.42 & 0.20 & 0.38 &  0.40 & 0.38 \\ 
%\model (500) &  0.31 &  0.42 & 0.19 & 0.37 &  0.39 & 0.37 \\

%\model (4000) alt.summary & 0.30 & 0.39 & 0.20 & 0.39 & 0.39 \\ 
%\model (4000) para.summary & 0.30  &  0.39 &   0.21  &  0.40 &  0.39 \\ 
\hline
%TSSP+CSSL&  0.85 & 0.10  & - &  -  & - & - \\
%\rowcolor{verylightgray} 
\multicolumn{7}{c}{Wikisection dataset ($k=6$)} \\
\hline
(8)  & CATS~\cite{somasundaran2020two} &  - & - & \textbf{0.113}  & - & - \\
(9) & Gen (seg+label)~\cite{inan2022structured} &    8000 & \xmark/\xmark & 0.188  & \textbf{0.873} & \textbf{0.925} \\
(10)   & \model &   7000+1000 & \cmark/\cmark & 0.134   & 0.866 &  0.924 \\
\hline
%TSSP+CSSL&  0.57 & 0.82  & - &  -  & - & - \\
%\rowcolor{verylightgray} 
\multicolumn{7}{c}{QMSum dataset ($k=85$)} \\
\hline
(11) & CATS~\cite{somasundaran2020two} &  - & - & 0.469  & - & - \\
(12) & Gen (seg+label)~\cite{inan2022structured} &    8000 & \xmark/\xmark & \textbf{0.443}  & 0.196  & 0.326 \\
% 0.44    0.40    0.20    0.34    0.35    0.33 
(13) & \model &   7000+1000 & \cmark/\cmark & \textbf{0.443}   & \textbf{0.234} &  \textbf{0.365} \\
% 0.37    0.30    0.27    0.49    0.46    0.47 
% 0.28    0.28    0.22    0.35    0.36    0.35
\bottomrule
\end{tabular}
\label{tab:com-table}
\end{table*}

%\textbf{Q1: How does \model perform on conversational datasets?} Table~\ref{tab:com-table} shows that \model (row 4),  with global context (both static and dynamic context) enabled,  significantly outperforms the strongest baseline, Gen (seg+label) shown in row 3,  on podcast dataset according to the title metrics (paired t-test with p-value $<$ $0.05$).  A similar result is observed in the QMSum dataset (rows 14-15),  although statistically insignificant due to its small size (35 documents in the test set). This underscores the importance of capturing global context for generating high-quality  titles. We also observe that  the segmentation accuracy, as measured by WinDiff, remains close to the baseline,  suggesting that segmentation relies less on global context compared to title generation.

\textbf{Q1: How does \model perform on conversational datasets?}
Table~\ref{tab:com-table} shows that \model (row $4$), with both static and dynamic context enabled, significantly outperforms the strongest baseline, Gen (seg+label) (row $3$), on the podcast dataset according to title metrics (paired t-test, p-value $<$ $0.05$). A similar trend is observed in the QMSum dataset (rows $11$-$13$), though not statistically significant which might be due to its small test set ($35$ documents). This highlights the importance of capturing global context for high-quality title generation. Segmentation accuracy, measured by WinDiff, remains close to the baseline, indicating that segmentation relies less on global context.
Comparison with CATS (row $1$) suggests that coherence modeling in this method is less effective on conversational datasets compared to structured texts. The lower performance of GPT-4 zero-shot inference (row $2$) highlights the challenge of chapterizing long conversational documents without fine-tuning, even for powerful models like GPT-4. On Wikisection (rows $8$-$10$), where documents are short and well-structured, our model performs comparably to Gen(seg+label), as expected.

\textbf{Q2: Do static and dynamic context contribute equally to improving title metrics?}
The results in Q1 suggest that our new contextual features improve the title quality of conversational data more than boundary accuracy. To examine the individual effects of static and dynamic context on titles, we conduct an ablation study (rows $5$-$7$). Disabling static context (row $6$) causes a more significant decrease in title metrics than disabling dynamic context 
(row $5$).\footnote{Although enabling only the dynamic context (row $6$) improves boundary metrics compared to when both features are disabled (row $7$), but slightly reduces title quality.}After examining a few examples, we speculate that lower performance in the dynamic context-only model may be due to a chapterization style different from the ground truth,\footnote{For example, in an episode about scary stories, the model using dynamic context predicted generic titles like \textit{``Story 1’’}, \textit{``Story 2’’\textit}, and so on, , whereas the ground truth had specific titles like (\textit{``Invisible Humanoid''}, \textit{``Family of Sasquatch''}, …).} 
hinting at the insufficiency of the state-of-the-art reference-based metrics and a single ground truth for chapterization.

For a deeper understanding of the context's effect on titles, particularly \textit{title consistency} across chapters within an episode, we examine title length variation. We compute the coefficient of variation\footnote{Coefficient of variation = (title length std) / (mean title length).} for each episode and average it across the test set. Higher average coefficients indicate lower consistency. 
The baseline (row $3$) shows the highest variation (0.6), while \model and the dynamic context-only model score the lowest (0.55). The static context-only model’s score (0.58) is close to the baseline. These results highlight the limitations of reference-based metrics used in Table~\ref{tab:com-table} and show that dynamic context positively contributes to title quality, aligning with the original motivation for this feature (conditioning the next title on the already predicted ones).

\textbf{Q3: Do longer documents benefit more from global context?}
The primary rationale behind integrating global (static and dynamic) context in \model's input was to improve the chapterization of long documents that exceed the model’s context size. Thus, we hypothesized that longer documents would benefit more from \model compared to the baselines. This hypothesis is validated by the findings in Table~\ref{tab:long-short}. The first row shows the percentage improvement in title quality over the baseline, Gen(seg+label), for documents fully processed by \model. The second row shows improvements for longer documents requiring chunking, which make up 80\% of the test data. It is evident that longer documents see more substantial improvements. Table~\ref{tab:anecdotal-examples} demonstrates how using metadata for chapterizing long documents enhances chapter titles’ informativeness. \model adds words like ``Planet" and ``Sandeep" from the metadata, which are missing in the input chunk with chapter boundaries due to an already established context.

\textbf{Q4: How does the length and source of the static context impact chapterization?}
%Since podcast creators provide both chapters and metadata (title and description) 
%for their episodes, it is natural to assume that episode descriptions \textit{may} resemble 
%chapter titles. This motivated us to explore whether having metadata 
%similar to chapter titles is essential for better chapterization. We defined similarity as the 
%percentage of chapter titles’ vocabulary present in the metadata and computed the Spearman 
%correlation between this and the ROUGEL$_{F1}$ score. We observed a negligible positive 
%correlation ($0.08$), indicating the LLMs' ability to generalize reasoning.
To test if longer static context enhances auto-generated chapter quality, 
we computed the Spearman rank correlation between static 
context length and the $\Delta$ROUGEL$_{F1}$ of 
\model with and without static context. 
We found a negligible negative correlation, suggesting 
that longer static context does not necessarily improve metric scores.

Given the increasing use of LLMs for content generation, 
we further explored the robustness of \model to AI-generated static 
context.
For this, we instructed GPT-4 to generate episode descriptions 
based on the episode transcripts and used them in place of creator-provided 
descriptions. As a result,
we observed a $7\%$ drop in ROUGEL$_{F1}$ compared to 
\model that uses creator descriptions (row $4$ in Table~\ref{tab:com-table}). We conclude that creator-provided static context is more effective for chapterization.

\begin{table}[ht]
%\small
\footnotesize
\caption{\model's title metrics improvement (\%) over the baseline Gen (seg+label) for short (no chunking needed) and long (chunking needed) transcripts.
%(no chunking needed) and long transcripts (chunking needed).
}
\label{tab:long-short}
\begin{tabular}{P{1.5cm}ccP{2cm}}
\toprule
%Annotator  & \multicolumn{1}{l}{nDCG} & \multicolumn{1}{l}{R@30} & \multicolumn{1}{l}{R@50} & \multicolumn{1}{l}{R@100} & \multicolumn{1}{l}{RR} \\ \midrule

Needs chunking? & $\Delta$ROUGEL$_{F1}\%$ & $\Delta$SBERT$_{F1}\%$ &  $\%$ in Test data\\
\midrule
NO & 6.61 & 2.92 & 20\\
YES & 12.08 & 5.51 & 80 \\
\bottomrule
\end{tabular}
\end{table}

\begin{table}[ht]
%\small
\footnotesize
%\centering
\caption{Anecdotal examples indicating how metadata can improve informativeness of 
chapter titles in \model in comparison with the baseline. 
Gen (seg+label) is unable to infer the bold words from 
its input text.
}
\label{tab:anecdotal-examples}
\begin{tabular}{P{2.5cm}P{2.5cm}P{2cm}}
\toprule
%Annotator  & \multicolumn{1}{l}{nDCG} & \multicolumn{1}{l}{R@30} & \multicolumn{1}{l}{R@50} & \multicolumn{1}{l}{R@100} & \multicolumn{1}{l}{RR} \\ \midrule
\model & Ground-truth & Gen (seg+label)\\
\midrule
% Most Anticipated Movies of 2020	Most anticipated movies of 2020	The Movies
Sigma MC\textbf{\textcolor{orange}{-11}} Adapter & Gear of the Week: Sigma’s MC\textbf{\textcolor{orange}{-11}} & Sigma Mount Converter\\
\midrule
\textbf{\textcolor{orange}{Planet}} of Lana Reviews  & \textbf{\textcolor{orange}{Planet}}  of Lana Reviews & \vfill Lana Review\\
\midrule
\textbf{\textcolor{orange}{Sandeep}}'s Family Background &	About \textbf{\textcolor{orange}{Sandeep}}'s Background	& \vfill Birth \\
\bottomrule
\end{tabular}
\vspace{-0.4cm}
\end{table}

\section{Deployment}\label{sec:deployment}

Podcast chapters with creator-provided timestamps have been available on our platform. There overall coverage, 
however, is low.
In April 2024, we started a limited roll-out of our chapterization model. % to 90\% of users. 
Since auto-generated chapters broadens availability of chapters, 
we expect that if they have high quality, we would see higher engagement with chapters. 
Overall we saw an $88.12\%$ increase in chapter-initiated plays after the roll-out.

\begin{figure}[t]
\centering
    \includegraphics[scale=0.4]{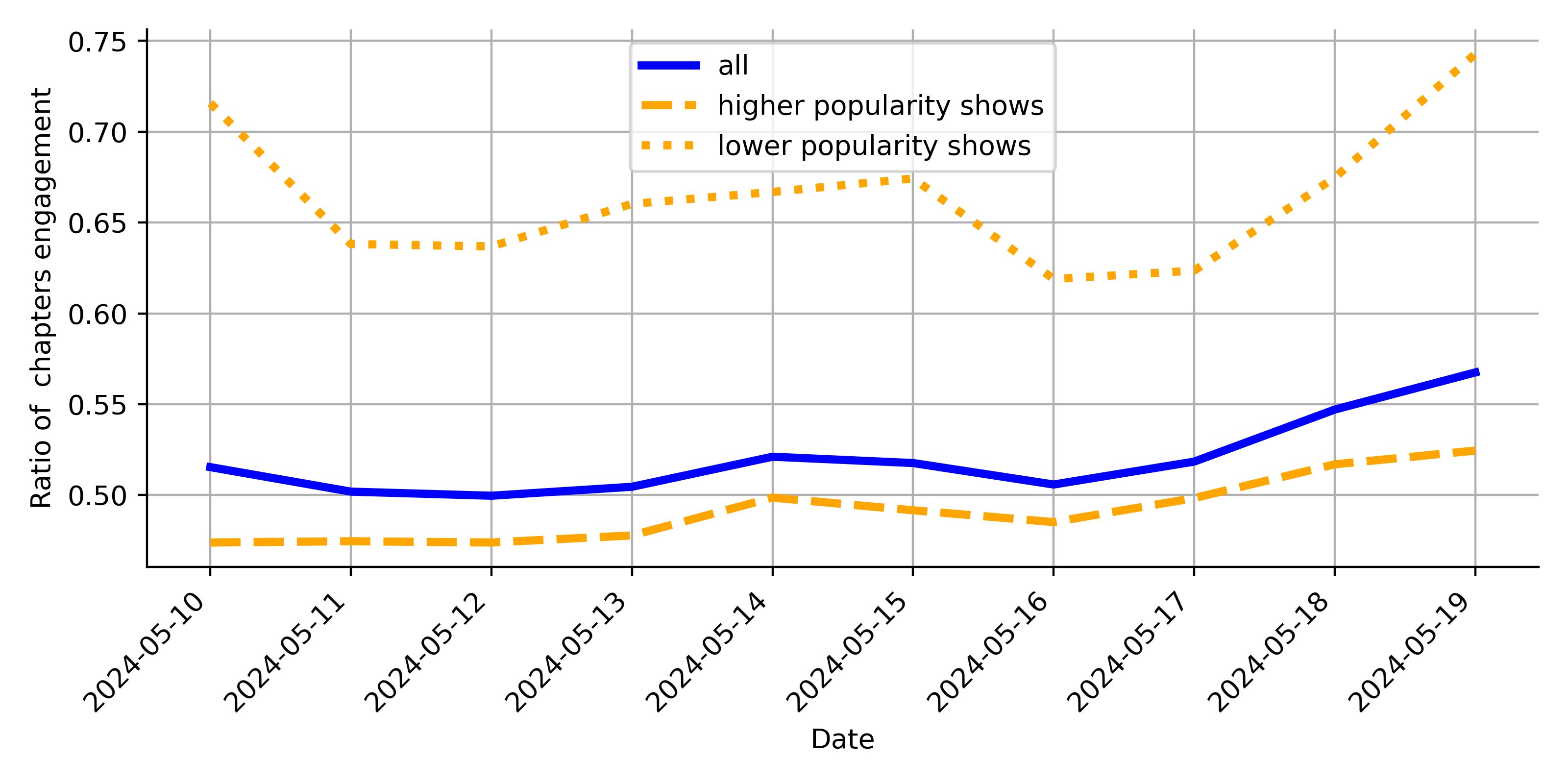}  
    \caption{Ratio of relative chapters engagement between episodes with auto-generated titles and creator-provided titles, plotted as the moving average over previous 7 days.}
    \vspace{-0.3cm}
    \Description{Ratio of relative chapters engagement between episodes with auto-generated titles and creator-provided titles, plotted as the moving average over previous 7 days.}
    \label{fig:ratio_mov_ave}
\end{figure}

To understand user engagement with podcast chapters, we measured the percentage of listeners who interacted with chapters by 
playing or scrolling them.
% scrolling or clicking. 
We compared engagement ratios between episodes with auto-generated chapters and those with creator-provided chapters. A lower ratio would indicate that auto-generated chapters are less attractive or useful. Our model, designed to mimic creator chapters, assumes this ratio should not exceed 1. We collected data over the last $10$ days and plotted the $7$-day moving average in Figure~\ref{fig:ratio_mov_ave}. The overall trend shows stable, positive growth. Notably, less popular shows had a higher engagement ratio (almost $0.75$) compared to more popular shows (about $0.53$). This suggests that auto-generated chapters are particularly beneficial for less popular shows, enhancing user engagement effectively.

%\textcolor{red}{K: removed the unique users analysis cause nobondy likes/understands it}
%\begin{figure}[t]
%\centering
%        \includegraphics[scale=0.45]{images/num_unique_users.png}  
%    \caption{Distribution of number of unique episode users for episodes published since the deployment of the chapterization model.\label{fig:uniq_users}}
%\end{figure}

\begin{figure}[ht]
\centering
    \centering
    \includegraphics[width=0.85\columnwidth]{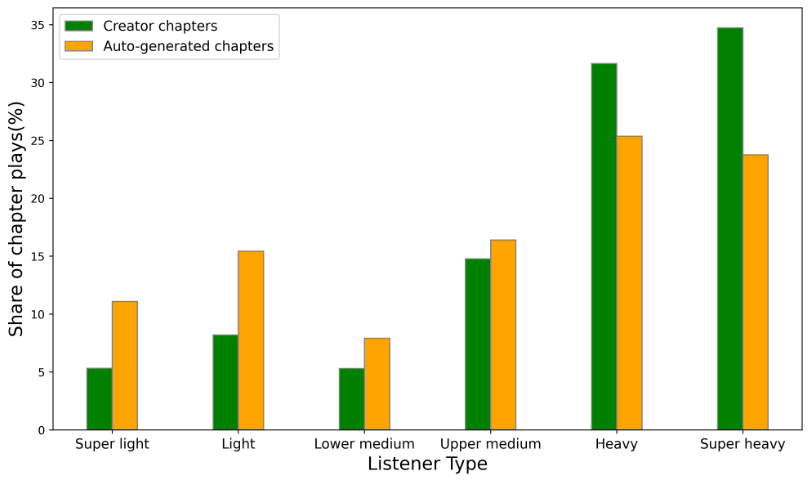}
    \caption{Percentage of creator-provided and auto-generated chapter plays across five 
    user groups based on consumption.}
    \Description{Percentage of creator-provided and auto-generated chapter plays across five 
    user groups based on consumption.}
    \label{fig:listener-type}
    \vspace{-0.3cm}
\end{figure}

We divided users into five groups based on their total consumption since \model's deployment and calculated the percentage of chapter plays for each group.  Figure~\ref{fig:listener-type} shows that creator-provided chapters are predominantly used by heavy listeners, likely due to their lower coverage. In contrast, auto-generated chapters have a more balanced distribution, with $50.84\%$ of play counts from super light to upper medium users. 
%This indicates  
%that auto-generated chapters help users with limited time navigate episode content efficiently.
This shows auto-generated chapters help users with limited time navigate episodes efficiently.
%suggesting chapters help users with limited time navigate content efficiently. 
%Comparing distributions, we observe that episodes with auto-generated chapters have a larger share of heavy listeners and fewer super light users. The broader coverage of auto-generated chapters likely explains this pattern.
%\textcolor{red}{compare CC and ML?}

%To examine the impact of episode duration on chapter usage,  we computed the Spearman correlation between duration of episodes with auto-generated  chapters and  the share of streamers who play chapters.  We found a weak correlation ($0.17$),  indicating that duration alone does not determine chapter usage. Notably, in entertainment categories like ``TV and Shows'', ``Leisure'', and ``Arts'',  longer episodes receive more chapter plays,  showing that duration and the episode content together can influence chapter usage.

To examine the impact of episode duration on chapter usage, we computed the Spearman rank correlation between duration of episodes with auto-generated 
chapters and the percentage of their chapter users.
%\textcolor{red}{I don't get this: the share of streamers who play chapters}. 
The weak correlation ($0.17$) indicates that episode duration alone does not determine chapter usage. However, in entertainment categories like ``TV and Shows'', ``Leisure'', and ``Arts'', longer episodes receive more chapter plays, suggesting that both duration and content influence chapter usage.

\section{Extrinsic Evaluation}\label{sec:extrinsic_evaluation}
%Textual descriptions of podcast episodes may not entirely capture the details or relevant content that listeners are looking for. These details are typically found within transcripts, which are often notably long, and thus
%The sparse retrieval based on BM25 is known to struggle with long
%documents, potentially failing to properly reward the occurrence of a
%query term in such long documents~\cite{lv2011documents}. 
%indexing them for sparse retrieval would incur
%substantial storage cost.
%\textcolor{red}{report a number}.
%Therefore, we propose to leverage chapter titles as readily available episode summaries in place of the full transcripts and  augment them to episode descriptions. We estimate that this would save cost by at least $10$ times\footnote{We compare the  size of the inverted index created by indexing chapter titles with that generated from  the whole transcripts.} compared to indexing the entire transcripts. We hypothesize that augmenting podcast descriptions with chapter titles  significantly enhances the effectiveness of sparse retrieval in search  since they contain important terms that users might be searching for. 

%\textcolor{red}{The main motivation for this work is that most of our podcasts are not chapterized. Previously, we focused on chapterization for navigation, but it can also enhance other tasks, such as the one discussed here.}
Podcast chapterization primarily aims at facilitating navigation through episode content. 
%In this section, we demonstrate how podcast chapters can enhance episode search retrieval as 
%a downstream task.
This section shows how podcast chapters can also enhance episode search retrieval as 
a downstream task.
%\textcolor{red}{Does it look good?}

Textual descriptions of podcast episodes often miss key details that listeners seek. These details are usually in the transcripts, which are lengthy and costly to index. We propose using chapter titles as summaries instead of full transcripts to enhance episode descriptions. This approach could reduce costs by at least tenfold\footnote{We compare the  size of the inverted index created by indexing chapter titles with that generated from 
the whole transcripts.}  compared to indexing entire transcripts. We believe that adding chapter titles to descriptions will significantly improve sparse retrieval in search by including important terms users search for.

To test this hypothesis, we design an experiment to explore the impact of  indexing chapter titles on search effectiveness. For this, we use the  TREC podcast dataset~\cite{jones2021trec} collected for short  segment retrieval and summarization task. This dataset contains  human relevance judgments for $54$ search queries, 
of $3$ types: topical, re-finding, and known items. 
a pool of ~$100$k episodes, and ~$900$ labeled query-episode pairs. Note that in this experiment, we perform retrieval and report metrics on episode-level and not segment-level.
We use BM25, implemented by Anseri~\cite{yang2018anserini}, as the retrieval method, measure search success by nDCG, recall, and Reciprocal Rank (RR), and consider $4$ methods for indexing episodes:
\footnote{We leave the efficient application of the 
costly document expansion methods based on abstractive 
summarization~\cite{jeong2021unsupervised, 
pan2024semantically} or Doc2Query~\cite{nogueira2019document, 
nogueira2019doc2query} as future work.} 
\squishlist
    \item \textbf{Desc}: Only episode descriptions are indexed.
    \item \textbf{Desc+princ}: Descriptions are expanded with key sentences of the transcripts, extracted using the Principal Uniq-Ind~\cite{zhang2020pegasus}.\footnote{This method computes the 
    ROUGE1$_{F1}$ score between each sentence $s_i$ and rest of the transcript, selecting the top sentences with the highest scores. ``Uniq'' means only unique $n$-grams are considered and ``Ind" indicates independent scoring of sentences. Extractive summaries are limited to $24$ words for fair comparison with chapter titles.}
    \item \textbf{Desc+chap}: Descriptions are expanded with chapter titles and then indexed.
    \item \textbf{Desc+trans}: Both descriptions and full transcripts are indexed. This is expected to perform the best despite the high cost.
\squishend

Table~\ref{tab:extrinsic_eval} summarizes the results. 
We observe that \textbf{Desc+chap} significantly outperforms 
the baselines (\textbf{Desc} and \textbf{Desc+princ}) 
according to nDCG, R@30, and R@50. This demonstrates that chapter titles effectively capture the essence of the transcript while significantly reducing the storage needed for indexing.

\begin{table}[t]
\footnotesize
\caption{Extrinsic results for the TREC podcast dataset. The $^\ddagger$ denotes statistical significance when compared to the Desc+princ using Students' t-tests at 0.95 confidence interval. 
While Desc+trans is more effective %as it uses entire transcripts, 
its index size is more than 10 times bigger than Desc+chap.
}
\label{tab:extrinsic_eval}
\begin{tabular}{@{}p{1.5cm}lllll@{}}
\toprule
Setting  & \multicolumn{1}{l}{nDCG} & \multicolumn{1}{l}{R@30} & \multicolumn{1}{l}{R@50} & \multicolumn{1}{l}{R@100} & \multicolumn{1}{l}{RR} \\ \midrule
Desc & 0.239 & 0.241 & 0.264 & 0.324 & 0.441 \\
Desc+princ & 0.243 & 0.242 & 0.254 & 0.322 & 0.440 \\
Desc+chap & 0.276$^\ddagger$ & 0.265$^\ddagger$ & 0.315$^\ddagger$ & 0.362 & 0.528$^\ddagger$ \\ 
\cline{1-6}
Desc+trans & 0.336 & 0.374 & 0.422 & 0.516 & 0.446\\
\bottomrule
\end{tabular}
\vspace{-0.4cm}
\end{table}

% \begin{table}[] 
%     \caption{Caption}
%     \centering
%     \begin{tabular}{lccccc}
%         \toprule
%          Setting name & nDCG@$30$ & R@$30$ & R@$50$ & R@$100$ & MRR\\
%          \midrule
%          Only-desc & $0.239$ & $0.241$ &  $0.264$ & $0.324$ & $0.441$ \\
%          Desc+trans & $0.247$ & $0.248$ &  $0.272$ & $0.344$ & $0.453$ \\ 
%          Desc+principal & $?$ & $?$ &  $?$ & $?$ & $?$\\
%          Desc+chap & $0.276$ & $0.265$ &  $0.315$ & $0.362$ & $0.528$ \\ 
%          \bottomrule
%     \end{tabular}
%     \label{tab:extrinsic_eval}
% \end{table}
%The podcast track was designed to encourage research into podcasts in the information retrieval and NLP research communities. The track consisted of two shared tasks: segment retrieval and summarization, both based on a dataset of over 100,000 podcast episodes (metadata, audio, and automatic transcripts). The segment retrieval task aims at finding the most relevant episode segments as judged by NIST assessors for a given set of queries. These queries were particularly designed for targeting episode segments of 2-mins duration. There are three types of queries in this dataset: topical, refinding, and known items. We use all the ~50 queries in our evaluation.  Note that in offline evaluations, it is typical to have this many queries. Some example queries from this dataset are: “coronavirus spread”, “story about riding a bird”, “hvac industry environmentalism”, “racism in canada”, “cost of childcare”.
\section{Conclusion}\label{sec:conclusion}
%We presented a chapterization model that can efficiently chapterize 
%podcast episodes at scale using small LLMs. Our model allows for capturing long-range dependency in LLMs with limited context window by incorporating short global  context in each transcript chunk. We evaluate our model  on internal and public datasets and demonstrate its  competitive performance on both structured and conversational data.  We recently deployed the model on our platform and  observed that users find auto-generated chapters helpful  for browsing through  the episodes' content. We further show that chapter titles provide concise and  informative summary of the transcript that can enrich  episode descriptions for improving search effectiveness. 

We developed a chapterization model that efficiently processes podcast episodes at scale using small LLMs. Our model captures long-range dependencies by incorporating short global context in each transcript chunk. We evaluated our model on internal and public datasets, demonstrating its competitive performance on both structured and conversational data. After deploying the model on our platform, we observed that users find auto-generated chapters helpful for browsing episode content. We also showed that chapter titles provide concise and informative summaries of transcripts, enhancing episode descriptions and improving search effectiveness.

We acknowledge that podcast chapterization is subjective, and a single ground-truth reference may not fully capture the model’s capabilities. Therefore, we plan to extend our evaluation to include reference-free metrics. Additionally, we aim to leverage other modalities, such as audio and video, to further improve chapterization.

%We acknowledge that podcast chapterization  is a subjective task, and thus a reference-based evaluation  based on a single ground-truth  may not properly capture the model's capabilities. Therefore, we plan to extend our evaluation  framework to reference-free metrics.  Another future direction is to leverage other modalities (audio and video) for  improving chapterization.

%\clearpage
\bibliographystyle{ACM-Reference-Format}
\bibliography{fp-2024-chapterization}

\end{document}